\documentclass[9pt, conference]{IEEEtran}

\usepackage{cite}
\usepackage{amsmath,amssymb,amsfonts}
\usepackage{graphicx}
\usepackage{textcomp}
\usepackage{xcolor}

\usepackage{subcaption}

\usepackage{multirow}
\usepackage{diagbox}

\usepackage{pifont} 
\newcommand{\cmark}{\textcolor{green}{\ding{51}}}
\newcommand{\xmark}{\textcolor{red}{\ding{55}}}

\usepackage{algorithm}
\usepackage{algorithmicx}
\usepackage{algpseudocode}

\usepackage{fvextra}

\usepackage[altpo]{backnaur}

\usepackage{xspace}
\usepackage{hyperref}

\newcommand{\sys}{\textsc{Flag}\xspace}

\begin{document}

\title{\sys: Formal and LLM-assisted SVA Generation for Formal Specifications of On-Chip Communication Protocols}

\author{
\IEEEauthorblockN{Yu-An Shih, Annie Lin, Aarti Gupta and Sharad Malik}
\IEEEauthorblockA{
Princeton University, Princeton, USA \\
\{yashih, al2413\}@princeton.edu, aarti@cs.princeton.edu, sharad@princeton.edu}
}

\maketitle

\begin{abstract}
Formal specifications of on-chip communication protocols are crucial for system-on-chip (SoC) design and verification. However, manually constructing these formal specifications from informal documents remains a tedious and error-prone task. Although recent efforts have used Large Language Models (LLMs) to generate SystemVerilog Assertion (SVA) properties from design documents for Register-Transfer Level (RTL) design verification, in our experience these approaches have not shown promise in generating SVA properties for communication protocols. Since protocol specification documents are unstructured and ambiguous in nature, LLMs often fail to extract the necessary information and end up generating irrelevant or even incorrect properties.

We propose \sys, a two-stage framework to help construct formal protocol specifications from informal documents. In the first stage, a predefined template set is used to generate candidate SVA properties. To avoid missing necessary properties, we develop a grammar-based approach to generate comprehensive template sets that capture critical signal behaviors for various communication protocols. In the second stage, we utilize unambiguous timing diagrams in conjunction with textual descriptions from the specification documents to filter out incorrect properties. A formal approach is first implemented to check the candidate properties and filter out those inconsistent with the timing diagrams. An LLM is then consulted to further remove incorrect properties with respect to the textual description, obtaining the final property set. Experiments on various open-source communication protocols demonstrate the effectiveness of \sys in generating SVA properties from informal documents.
\end{abstract}


\section{Introduction}

Modern system-on-chips (SoCs) rely on standardized on-chip communication protocols (e.g., AXI \cite{spec:axi}, WISHBONE \cite{spec:wishbone} and PCI \cite{spec:pci}) to regulate the interaction between various hardware intellectual properties (IPs). These protocols define the rules and mechanisms for data exchange between multiple hardware components, ensuring efficient and reliable communication between processors, memory, and peripherals.

As a consequence, formal specifications of these protocols play a crucial role in SoC design and verification \cite{roychoudhury2003using, lin2008amba, chen2010synthesizable, spec:axichecker}. A formal specification consists of a set of temporal properties that regulate the behavior of interface signals between multiple hardware components. These properties, typically written in SystemVerilog Assertions (SVA) \cite{10458102}, are used to debug hardware components through assertion-based verification (ABV) \cite{witharana2022survey}. In addition to hardware design verification, these formal specifications are used in various electronic design automation (EDA) tasks, including hardware synthesis \cite{bloem2007automatic, godhal2013synthesis}.

Protocol specification documents are often written informally in natural languages to assist hardware design engineers in developing systems and modules compatible with the protocol. To construct a formal specification, verification engineers have to read the document, extract relevant information about signal interaction rules, and translate them into formal properties \cite{roychoudhury2003using}. This process can be tedious and error-prone. Moreover, each time the protocol is updated to a new version, the specification has to be modified accordingly. Therefore, there is generally a lack of open-source formal protocol specifications despite the wide adoption of ABV.

Recent research has explored the use of machine learning (ML), including Large Language Models (LLMs), to generate SVA properties from IC design specification documents \cite{parthasarathy2021spectosva, mali2024chiraag, yan2025assertllm}. The primary goal of these works is to support RTL design and verification. ML models are consulted to extract information from documents and translate them into SVA properties. The results are evaluated using metrics such as number of generated assertions and cone-of-influence (COI) coverage. However, in our experience, the problem of generating formal specifications for communication protocols brings new challenges that are not solved by these approaches.

First, a formal protocol specification requires the exact information that describes how interactions between signals are regulated. As specification documents typically consist of hundreds of pages, we find that it is very difficult for ML models to understand which information to extract. If applied directly, they end up missing necessary properties while generating many irrelevant ones.

Second, statements in informal protocol specifications are meant to be read by experienced human engineers. Some of these can be ambiguous or difficult to understand for ML models. As shown in Fig.~\ref{fig:ambiguous}, it can be intuitive to translate an ambiguous natural language statement into an incorrect SVA property. In fact, ML models in previous works \cite{keszocze2019chatbot, aditi2022hybrid} have made the exact type of mistake shown in Fig.~\ref{fig:ambiguous}, when translating natural language sentences into SVAs.

\begin{figure}
    \begin{Verbatim}[breaklines=true, breaksymbolleft={}, commandchars=\\\{\}, fontsize=\small]
\textbf{Natural Language statement}
    DATA must remain stable when VALID is asserted and READY is LOW.
\textbf{Incorrect SVA property}
    (VALID && !READY) |-> $stable(DATA)
\textbf{Correct SVA property}
    (VALID && !READY) |-> ##1 $stable(DATA)
    \end{Verbatim}
    \captionsetup{type=figure}
    \caption{Example of an ambiguous statement. An ML model may intuitively translate ``remain stable" into the \$stable operator in SVA. However, the ``remain stable" in this context actually means ``assign the same value in the next clock cycle," i.e., the \$stable operator should be delayed by one cycle.}
    \label{fig:ambiguous}
\end{figure}

\begin{figure*}
    \centering
    \includegraphics[width=\textwidth]{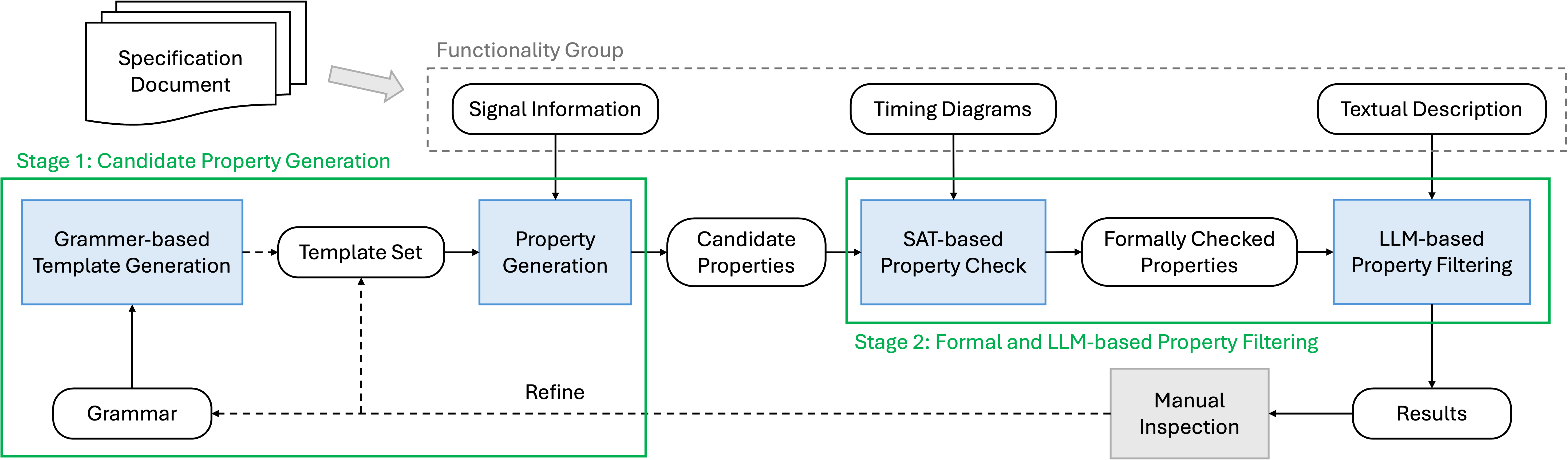}
    \caption{The \sys framework.}
    \label{fig:overview}
\end{figure*}

To address these issues, we propose \sys (\underline{F}ormal and \underline{L}LM-assisted \underline{A}ssertion \underline{G}eneration), a two-stage framework to assist its user in generating formal protocol specifications from informal documents (Fig.~\ref{fig:overview}). In addition to LLM-based techniques, \sys combines a template-based property generation method and a formal property check algorithm to obtain accurate SVA properties.

In the first stage, \sys generates a set of candidate properties based on a predefined template set. A grammar-based approach is developed to generate the template set, where the grammar is constructed by studying signal behaviors of various on-chip communication protocols. Compared to ML-based methods, \emph{this approach avoids missing necessary properties or generating irrelevant ones}. Another advantage of the template-based approach is that it guarantees 100\% syntax correctness and improves readability, while LLMs can use complex syntax for simple properties or even make syntax errors.

\begin{figure}
  \centering
  \begin{subfigure}[b]{0.48\columnwidth}
      \centering
      \includegraphics[width=\textwidth]{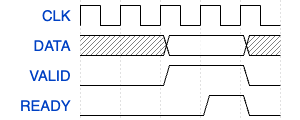}
      \caption{VALID before READY.}
      \label{fig:td-1}
  \end{subfigure}
  \begin{subfigure}[b]{0.48\columnwidth}
      \centering
      \includegraphics[width=\textwidth]{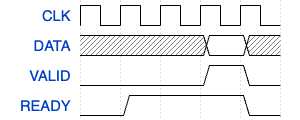}
      \caption{READY before VALID.}
  \end{subfigure}
  \caption{Example of timing diagrams extracted from the AMBA AXI Protocol Specification~\cite{spec:axi}.}
  \label{fig:tds}
\end{figure}

In the second stage, \sys utilizes timing diagrams in the specification documents to \emph{formally check} the candidate properties. Unlike natural language descriptions, these timing diagrams describe signal behavior in a cycle-by-cycle manner and are thus unambiguous in nature. For example, one can easily distinguish the correct and incorrect SVA properties depicted in Fig.~\ref{fig:ambiguous} by observing the timing diagrams in Fig.~\ref{fig:tds}. By filtering out properties that are inconsistent with the timing diagrams, we reduce the chance of obtaining incorrect properties due to text ambiguity. However, note that some incorrect properties may still be consistent with the descriptive timing diagrams in the document.

Therefore, given the formally checked properties, we prompt an LLM to further filter out incorrect properties with respect to a textual description extracted from the specification document. The key concept is that LLMs are now used to compare the textual descriptions with a set of formally generated properties, instead of being asked to generate SVA properties provided only the massive and unstructured documents. With this approach, we mitigate the risks of missing necessary properties and generating incorrect or irrelevant properties from ML models.

Our contributions are summarized as follows:
\begin{itemize}
    \item We propose \sys, a template-based framework that combines formal and LLM-based techniques to help generate formal protocol specifications from informal documents.
    \item We develop a grammar-based approach to generate and maintain template sets for the \sys framework, and construct a grammar that can be used to generate SVAs for various on-chip communication protocols.
    \item The effectiveness of \sys is evaluated on various real-world communication protocols. This includes a comparison with a state-of-the-art approach for SVA generation from design specification documents.
\end{itemize}
We publicly release our code and experiments at \url{https://github.com/Yu-An-Shih/FLAG}.

The remainder of this paper is organized as follows. Section~\ref{sec:overview} describes the overview of our approach. Section~\ref{sec:generation} and \ref{sec:filtering} explain the details of our algorithms. In Section~\ref{sec:experiment}, we evaluate each step of our algorithms, and compare our results with a state-of-the-art SVA generation framework. Section~\ref{sec:limitations} indicates limitations of our framework, and Section~\ref{sec:related} discusses related works. Finally, we provide concluding remarks and indicate future directions in Section~\ref{sec:conclusion}.
\section{Overview} \label{sec:overview}

The overall workflow of our \sys framework is shown in Fig.~\ref{fig:overview}. This section provides an overview of each step to describe how the framework assists its users in generating SVA properties from specification documents.

\textbf{User Input.} To begin with, the user extracts a group of timing diagrams from the specification documents. The group of timing diagrams is used to describe a specific functionality of the protocol. For example, we extracted timing diagrams for 3 functionality groups from the AMBA AXI Protocol Specification \cite{spec:axi}: the master/slave channel handshake process, the reset operation, and the low-power request/acknowledge handshake mechanism. The timing diagrams of the master/slave channel handshake process are shown in Fig.~\ref{fig:tds}.

For each functionality group, the user indicates the relevant signals. The user is also required to extract textual descriptions that explain the target functionality from the documents. This can be a set of rules, a whole section, or even include signal descriptions.

\textbf{Property Generation.} Given relevant signal information, our property generation algorithm generates candidate properties based on a predefined set of templates. An example of such templates is shown in Fig.~\ref{fig:template}. The properties are generated by replacing the variables in the template with actual signals.

\begin{figure}[htbp]
    \begin{Verbatim}[breaklines=true, breaksymbolleft={}, commandchars=\\\{\}, fontsize=\small]
\textbf{Template}
    If [signal] is [level] and [signal] is [level], then [signal/word] remains stable in the next cycle.
\textbf{Candidate property 1}
    If VALID is HIGH and READY is LOW, then DATA remains stable in the next cycle.
\textbf{Candidate property 2}
    If VALID is LOW and READY is HIGH, then DATA remains stable in the next cycle.
\textbf{Candidate property 3}
    If VALID is LOW and READY is HIGH, then READY remains stable in the next cycle.
    \end{Verbatim}
    \captionsetup{type=figure}
    \caption{Example of a template and some of its candidate properties generated by the property generation algorithm.}
    \label{fig:template}
\end{figure}

\textbf{SAT-based Property Check.} The next step of our framework is to formally check whether the candidate properties hold on the timing diagrams. In this step, each temporal property, along with each timing diagram, will be translated into formulas in propositional logic and checked by Satisfiability (SAT) solvers. The details of how the check is modeled as a SAT problem are described in Section~\ref{sec:filtering:sat}.

\textbf{LLM-assisted Property Filtering.} Even if a property holds on all timing diagrams, it may still be incorrect with respect to the protocol. For example, \texttt{candidate property 3} in Fig.~\ref{fig:template} would pass the formal property check since it does not violate any timing diagrams in Fig.~\ref{fig:tds}, but it may not be a necessary constraint for a typical master/slave handshake mechanism. Therefore, after the SAT-based property check, an LLM is consulted to check whether the remaining properties are relevant and should hold with respect to the textual descriptions. The LLM serves as a proxy for human engineers, reading the documents and filtering out incorrect properties from an existing set.

\textbf{Template Set Generation.} Finally, the effectiveness of our algorithm depends on the quality of the predefined template set. We develop a grammar-based approach to generate compact yet comprehensive template sets, and explore how the grammar can be refined by studying results of generated formal specifications. The details are discussed in Section~\ref{sec:generation}.

Note that both the templates and properties are in abstract syntax tree (AST) format for the Property Generation and SAT-based Property Check tools to process. Our tool also translates them into formats such as natural language (as shown in Fig.~\ref{fig:template}) and SVA. The LLM reads the natural language format, and the SVA format can be used by formal verification tools for various applications.

\section{Template and Property Set Generation} \label{sec:generation}

As a template-based framework, \sys cannot generate a property if its structure is not included in the template set. On the other hand, if the template set is too large, the tool can end up generating millions of candidate properties, creating a burden on the checking algorithms. Therefore, a systematic approach is required to develop \emph{succinct yet comprehensive template sets}.

\subsection{Experience-based Template Set Development} \label{sec:generation:template}

At first glance, it may seem that an extremely large template set is required to cover all the SVA properties needed for various protocols. However, since communication protocols primarily regulate data transfer mechanisms, we find that a fairly simple set of structures is usually sufficient to cover most properties. For example, among all our experiments we have not seen a case requiring complex SVA operators such as non-consecutive repetition (\texttt{[=m:n]}). It is also unlikely that a complex chain of various operators such as \texttt{\$rose(sig1) \&\& \$fell(sig2) \&\& sig3 == val} would be needed.

Moreover, we observe that SVA properties required for various protocols share similar structures. For example, the template shown in Fig.~\ref{fig:template} is a typical structure for master/slave handshake mechanisms. This handshake mechanism appears in various communication protocols, including AXI, WISHBONE and PCI. Therefore, by studying property structures on previously solved cases, we can construct templates that are useful for other protocols.

Based on these observations, we propose an iterative approach to develop succinct yet comprehensive template sets. As Fig.~\ref{fig:overview} illustrates, the template set can be refined based on inspections of previous results to improve our framework in the future.

\subsection{Grammar-based Template Set Generation}

It can be challenging to maintain a template set as it gradually gets refined. For example, when adding a new template, the user would have to read the entire set to check whether it is redundant. If the set gets too large, it is also difficult to determine which templates to delete or modify. To solve these issues, we allow users to define \emph{grammars} that are more succinct and readable, compared to unstructured template sets.

A grammar consists of terminal symbols, non-terminal symbols, and a set of production rules (one for each non-terminal). In our framework, the terminals include (1) fixed SVA operators and (2) variables, including signal (\texttt{<signal>}, \texttt{<word>}) and value (\texttt{<level>}, \texttt{<value>}) types. 
The grammar can generate many \emph{sentences} over the terminal symbols by making some choice in each production rule, beginning with the top-level production rule.

In the example grammar shown in Fig.~\ref{fig:grammar}, the top-level rule defines an implication non-terminal, denoted $\langle$implic$\rangle$, which comprises an SVA implication operator (\textbf{\texttt{|->}}), a conjunction ($\langle$conj$\rangle$) non-terminal and a delay ($\langle$delay$\rangle$) non-terminal. Note that the conjunction rule has two choices in the grammar -- it can conjoin (\textbf{\texttt{\&\&}}) two or three assignments ($\langle$assign$\rangle$). Similarly, the delay rule has two choices -- it delays by one cycle (\textbf{\texttt{\#\#1}}) either an assignment or a stable non-terminal. The assignment rule equates (\textbf{\texttt{==}}) a signal and its level, and the stable rule comprises the stable operator (\textbf{\texttt{\$stable}}) on either a signal or a word.

In our grammars, each sentence generated by the top-level rule corresponds to a template. Note that each template has fixed SVA operators and the variables for signal and value types, where the variables will be filled in later during property generation. 
In particular, the template shown in Fig.~\ref{fig:template} can be generated by the grammar in Fig.~\ref{fig:grammar}.

\begin{figure}
    \begin{bnf*}
        \bnfprod{implic}
            {\textbf{\bnfts{|->}} \bnfsp \bnfpn{conj} \bnfsp \bnfpn{delay}}\\
        \bnfprod{conj}
            {\textbf{\bnfts{\&\&}} \bnfsp \bnfpn{assign} \bnfsp \bnfpn{assign} \bnfor}\\
        \bnfmore{\textbf{\bnfts{\&\&}} \bnfsp \bnfpn{assign} \bnfsp \bnfpn{assign} \bnfsp \bnfpn{assign}}\\
        \bnfprod{delay}
            {\textbf{\bnfts{\#\#1}} \bnfsp \bnfpn{assign} \bnfor \textbf{\bnfts{\#\#1}} \bnfsp \bnfpn{stable}}\\
        \bnfprod{assign}
            {\bnfts{\textbf{==} <signal> <level>}}\\
        \bnfprod{stable}
            {\bnfts{\textbf{\$stable} <signal/word>}}
    \end{bnf*}
    \captionsetup{type=figure}
    \caption{Example of a simple grammar, where the top-level rule is $\langle$implic$\rangle$, and the terminals are (1) SVA operators and (2) signal (\texttt{<signal>}, \texttt{<word>}) and value (\texttt{<level>}) types.}
    \label{fig:grammar}
\end{figure}

As shown in Fig.~\ref{fig:overview}, the grammars, as well as template sets, can be refined with the iterative approach by inspecting properties generated for previous protocols. Once a grammar is constructed, our tool \emph{automatically generates its corresponding template set}.

In Section~\ref{sec:experiment:grammar}, we demonstrate the effectiveness of this iterative refinement approach by constructing a single grammar for various communication protocols. However, note that it is also possible to derive specific grammars for targeted scenarios (e.g., deriving a specific grammar for the master/slave handshake mechanism).

\subsection{Template-based Property Generation}

Given a predefined template set, a straightforward algorithm is implemented to generate the corresponding candidate properties. The users are required to provide the signals and their corresponding types. Using Fig.~\ref{fig:template} as an example, \texttt{VALID} is a \texttt{signal} (one-bit signal) and \texttt{DATA} is a \texttt{word} (multi-bit signals). The property generation algorithm simply replaces each variable with actual signals according to its type. Note that in all our experiments, we only specify two signal types -- \texttt{signal} and \texttt{word}. However, it is possible to create more types to avoid generating incorrect properties (e.g. creating a \texttt{reset} type for reset signals).

\section{Formal and LLM-based Property Filtering} \label{sec:filtering}

Given a comprehensive set of candidate properties, the next step is to filter out the incorrect ones with respect to the protocol specification documents. First, a formal approach is used to check and remove properties that are inconsistent with the timing diagrams in the documents. We express both the properties and the timing diagrams as constraints in propositional logic, and then invoke a SAT solver to check their consistency. Next, LLM-based techniques are used to further remove incorrect properties with respect to a textual description extracted from the documents. This section describes the details of the filtering algorithms.

\subsection{SAT-based Property Check} \label{sec:filtering:sat}

\subsubsection{Encoding Timing Diagram Constraints}

Timing diagrams are used in specification documents to describe specific signal behaviors of the protocol. Such timing diagrams typically consist of signals with explicit, unconstrained, and symbolic values. For example, the timing diagram in Fig.~\ref{fig:td-1} can be expressed as follows: 
{\small
\begin{align*}
    & \texttt{DATA } = [\texttt{X, X, V1, V1, X}]\\
    & \texttt{VALID} = [\texttt{0, 0, 1, 1, 0}]\\
    & \texttt{READY} = [\texttt{0, 0, 1, 1, 0}]
\end{align*}
}
The signal vectors above can also be expressed as constraints on signal values at each clock cycle. As an example, the following constraint can be extracted from the above signal vectors:
{\small
\begin{equation}
\begin{split}
    \varphi_{td} \gets &(\texttt{DATA[2]} = \texttt{DATA[3]}) \wedge \\
                       &(\texttt{VALID[0]} = \texttt{0}) \wedge (\texttt{VALID[1]} = \texttt{0}) \wedge ...
\end{split}
\end{equation}
}
This formula defines a set of finite execution traces, which is implied by the timing diagram. Given the signal vector format, our tool automatically generates the constraints in propositional logic.

Note that timing diagrams in protocol specification documents are usually depicted in figure formats. They should be translated to the signal vector format for our tool to process. So far, we find it easy to perform this translation manually, since these descriptive timing diagrams typically contain only a few signals with fewer than 10 clock cycles. In addition, if some text format of these timing diagrams is provided (such as source codes to generate the figure format), it is possible to automatically translate it to the signal vector format.

We have also tried converting the timing diagrams from figure formats to signal vectors using LLMs. However, today's LLMs are still poor at analyzing the subtleties in figures. For example, LLMs often make mistakes when asked questions such as ``What is \texttt{VALID}'s value in each clock cycle?" for the timing diagrams in Fig.~\ref{fig:tds}.

\subsubsection{Encoding Property Constraints}

A temporal property enforces constraints to regulate signal behaviors on a target timing diagram, in addition to the constraints derived from the timing diagram itself.

For example, the following constraint can be derived from \texttt{candidate property 1} in Fig.~\ref{fig:template}:
{\small
\begin{equation}
\begin{split}
    \varphi_{p} \gets 
    & (\texttt{VALID[0]}\wedge\neg\texttt{READY[0]}\rightarrow\texttt{DATA[1]} = \texttt{DATA[0]}) \\
    & \wedge (\texttt{VALID[1]}\wedge\neg\texttt{READY[1]}\rightarrow\texttt{DATA[2]} = \texttt{DATA[1]}) \\
    & \wedge ...
\end{split}
\label{eq:prop}
\end{equation}
}
It can be seen that the temporal property is ``localized" to regulate signal behavior at each clock cycle in the timing diagram.

Algorithm~\ref{alg:encodeprop} provides a high-level view of our algorithm for encoding property constraints. The algorithm repeatedly calls a procedure \textsc{localize} to enforce the constraints of a candidate property on each clock cycle, and computes the conjunction of the returned propositional formulas.

\begin{algorithm}
\caption{Encoding property constraints} \label{alg:encodeprop}
\begin{algorithmic}[1]
    \Require Property (AST format) $prop$
    \Ensure Constraint (propositional logic) $\varphi_{p}$
    \State $\varphi_{p} \gets True$
    \For{$cyc \gets 0$ to $CYC\_MAX$}
        \State $\varphi_{local} \gets$ \Call{localize}{$prop, cyc$}
        \If{$\varphi_{local} \neq None$}
            \State $\varphi_{p} \gets \varphi_{p} \wedge \varphi_{local}$
        \EndIf
    \EndFor
\end{algorithmic}
\end{algorithm}

Since the property is provided in AST format, the \textsc{localize} function can recursively handle each operator, starting from the top-level operator, to generate the propositional constraint. For some examples, we show how the \textsc{localize} function handles the SVA operators of \texttt{candidate property 1} in Fig.~\ref{fig:template}:
\begin{itemize}
    \item Equality (\verb|==|): The equality operator has two operands, \texttt{signal} and \texttt{value}. The \textsc{localize} function returns $\texttt{signal}[cyc]=\texttt{value}$ for this operator.
    \item Stable (\verb|$stable|): This operator has a single \texttt{signal} operand. The formula $\texttt{signal}[cyc]=\texttt{signal}[cyc-1]$ would be returned. The \texttt{None} object (in Python) would be returned if $cyc=0$.
    \item Logical And (\verb|&&|): This operator has two operands, \texttt{lhs} and \texttt{rhs}. The formula $\textsc{localize}(\texttt{lhs}, cyc) \wedge \textsc{localize}(\texttt{rhs}, cyc)$ would be returned. However, the \texttt{None} object would be returned if either of the recursive calls returns \texttt{None}.
    \item Delay (\verb|##1|): The formula $\textsc{localize}(\texttt{operand}, cyc+1)$ is returned unless $cyc=CYC\_MAX$, where \texttt{None} would be returned.
    \item Implication (\verb+|->+): This operator has two operands, \texttt{ante} and \texttt{cons}. $\textsc{localize}(\texttt{ante}, cyc) \rightarrow \textsc{localize}(\texttt{cons}, cyc)$ is returned unless either of the recursive calls returns \texttt{None}, where \texttt{None} would be returned.
\end{itemize}
Note that equation (\ref{eq:prop}) can be derived by applying Algorithm~\ref{alg:encodeprop} to \texttt{candidate property 1} in Fig.~\ref{fig:template}.

\subsubsection{Checking Algorithm}

Finally, the overview of the SAT-based Property Check step is shown in Algorithm~\ref{alg:sat}. The \texttt{Encode\_td} and \texttt{Encode\_prop} functions extract the constraints enforced by each timing diagram and candidate property, respectively. Note that a property holds on a timing diagram if and only if it is impossible to find a valid execution trace on the timing diagram that violates the property. That is, $\varphi_{td} \wedge \neg \varphi_{p}$ is unsatisfiable. Therefore, a SAT solver can be used to check whether each candidate property is valid on the timing diagrams, retaining only the consistent ones.

\begin{algorithm}
    \caption{SAT-based Property Check} \label{alg:sat}
    \begin{algorithmic}[1]
        \Require Timing diagrams $TD$, candidate properties $CP$
        \Ensure Verified properties $VP$
        \ForAll{$td \in TD$}
            \State $VP \gets \phi$
            \State $\varphi_{td} \gets \text{Encode\_td($td$)}$
            \ForAll{$prop \in CP$}
                \State $\varphi_{p} \gets \text{Encode\_prop($prop$)}$
                \If{sat\_solve($\varphi_{td}\wedge\neg\varphi_{p}$) $= unsat$}
                    \State $VP \gets VP \cup \{prop\}$
                \EndIf
            \EndFor
            \State $CP \gets VP$
        \EndFor
    \end{algorithmic}
\end{algorithm}

\subsubsection{Optimizations}

Aside from the standard check flow, we perform further optimizations on the candidate properties. First, we remove all tautologies (e.g., ``If VALID is HIGH, then VALID is HIGH.") from the candidate set, since they do not add any constraint on signal behaviors. This can be done by checking the property with an entirely unconstrained timing diagram that contains only ``X" values. We perform a pre-process step to remove all tautologies before checking them with any provided timing diagrams.

In addition, we do not include vacuous properties in our check. That is, if the antecedent of an implication property never holds on any provided timing diagrams, we automatically remove it from the candidate set since its consequent can never be checked.

\subsection{LLM-based Property Filtering}

Given the formally checked property set, we further remove incorrect properties with respect to the textual description of the document. In this step, we provide the formally checked properties (in natural language) and textual description to an LLM, and prompt the model to \emph{extract} the properties from the given checked set that \textit{best describe the details} of the textual description. We use the LLM in a zero-shot setting -- without any task-specific fine-tuning or in-context examples -- relying solely on its pre-trained capabilities.

Due to the inherent stochasticity of LLMs, identical prompts may yield different outputs across multiple runs. To improve the robustness and reproducibility of our results, we issue the same prompt three times and take the union of the resulting sets of rules.

\section{Experimental Results} \label{sec:experiment}

We evaluate the effectiveness of \sys on six open-source on-chip communication protocols~\cite{spec:axi, spec:wishbone, spec:lowpower, spec:apb, spec:pci}, as shown in Table~\ref{tab:protocols}. For each protocol, we obtain timing diagrams from the specification documents and categorize them into different groups, based on the functionalities they describe. For each group, the relevant textual descriptions are also extracted from the documents. The number of timing diagrams used for each functionality group is shown in column~3, and the word count for each textual description is listed in column~4. Finally, we manually develop the required properties with respect to each functionality group that serve as the target set for evaluating the results, i.e., the ground truth for these experiments.

\begin{table}
\renewcommand{\arraystretch}{1.3}
\caption{Manually developed target properties for open-source on-chip communication protocols.}
\label{tab:protocols}
\centering
\begin{tabular}{|c|c|c|c|c|}
    \hline
    \multirow{2}{*}{\textbf{protocol}} & \multirow{2}{*}{\textbf{group}} & \multirow{2}{*}{\textbf{\# diagrams}} & \textbf{text size} & \multirow{2}{*}{\textbf{properties}} \\
    & & & \textbf{(\# words)} & \\
    \hline
    \multirow{3}{*}{AXI} & handshake & 3 & 779 & 2 \\
                         & reset     & 1 & 116 & 2 \\
                         & low power & 3 & 237 & 4 \\
    \hline
    \multirow{2}{*}{Q-Channel} & handshake & 2 & 584 & 11 \\
                               & reset     & 2 & 229 &  2 \\
    \hline
    \multirow{2}{*}{P-Channel} & handshake & 4 & 884 & 14 \\
                               & reset     & 3 & 654 &  2 \\
    \hline
    \multirow{2}{*}{WISHBONE} & data transfer & 6 & 1146 & 7 \\
                              & reset         & 1 &  608 & 2 \\
    \hline
    APB & data transfer & 6 & 439 & 12 \\
    \hline
    \multirow{2}{*}{PCI} & transactions & 2 & 1073 &  7 \\
                         & termination  & 8 &  753 & 20 \\
    \hline
\end{tabular}
\end{table}

The Z3 theorem prover \cite{deMoura2008} is used as the SAT solver for the SAT-based property check. For LLM-based property filtering, we select the OpenAI o1 reasoning model \cite{openai2024openaio1card}. All experiments were conducted on a dual-socket Intel Xeon Gold 6242 CPU (32 cores, 2.8GHz) with 32G RAM, using the OpenAI API to access the o1 reasoning model.

\subsection{Grammar-based Template Set Construction and Refinement}
\label{sec:experiment:grammar}

Table~\ref{tab:grammar} demonstrates how we construct and refine our grammar to provide a comprehensive template set that can generate temporal properties for various communication protocols.

We begin by constructing a grammar ($G1$) based on the PCI local bus protocol, since it contains the most complex data transaction mechanism among all our test cases. As shown in column~2, the template set generated by this PCI-based grammar covers all the properties required for APB and large portions for the other protocols. This supports our claim in Section~\ref{sec:generation:template} that properties required for different protocols often share similar structures.

After refining this grammar based on the Q-Channel protocol, the resulting grammar ($G2$) covers all protocols except WISHBONE, as shown in column~3. Finally, we obtained a complete grammar ($G3$) for all six protocols after refining it with WISHBONE.

\begin{table}
\renewcommand{\arraystretch}{1.3}
\caption{Grammar-based template set construction and refinement.}
\label{tab:grammar}
\centering
\begin{tabular}{|c|c|c|c|}
    \hline
    protocol & $G1$: PCI & $G2$: PCI + Q-Channel & $G3$: complete \\
    \hline
    AXI       &   7/8 (87.5\%) &   8/8 (100\%)  &   8/8 (100\%) \\
    Q-Channel & 10/13 (76.9\%) & 13/13 (100\%)  & 13/13 (100\%) \\
    P-Channel & 13/16 (81.3\%) & 16/16 (100\%)  & 16/16 (100\%) \\
    WISHBONE  &   7/9 (77.8\%) &   8/9 (88.9\%) &   9/9 (100\%) \\
    APB       & 12/12 (100\%)  & 12/12 (100\%)  & 12/12 (100\%) \\
    PCI       & 27/27 (100\%)  & 27/27 (100\%)  & 27/27 (100\%) \\
    \hline
\end{tabular}
\end{table}

In the next two sections, we evaluate our formal and LLM-based property filtering algorithms with the candidate property set generated from this complete grammar ($G3$).

\subsection{SAT-based Property Check}

Given the candidate properties generated from the complete grammar in Section~\ref{sec:experiment:grammar}, we implement the SAT-based property check to remove properties inconsistent with the timing diagrams.

Note that for some protocols (e.g. PCI), timing diagrams in different functionality groups regulate the behavior of the same signal sets. In these cases, the timing diagrams can be used together to filter candidate properties. For example, the PCI transactions case can use the timing diagrams in the termination case to further reduce incorrect properties from its candidate set.

The results are shown in Table~\ref{tab:sat}. The first two columns list the protocols and functionality groups. Column~3 shows the number of candidate properties generated for each functionality group. The number of properties retained after the SAT-based property check is shown in column~4. Column~5 shows the execution time of the algorithm (in seconds).

Since timing diagrams in protocol specification documents typically depict only a few signals with fewer than 10 clock cycles, the corresponding constraints (propositional formulas) usually contain only dozens of variables, which can be easily handled by modern SAT solvers. 
As shown in Table~\ref{tab:sat}, it only takes slightly more than one minute to check 11 thousand properties with 10 timing diagrams for the PCI termination case. Importantly, the formally checked sets retain only 5\% -- 22\% of the properties in the candidate sets in all cases. This benefits the next step by significantly reducing the input sizes to the LLM.

\begin{table}
\renewcommand{\arraystretch}{1.3}
\caption{SAT-based property check.}
\label{tab:sat}
\centering
\begin{tabular}{|c|c|c|c|c|}
    \hline
    \textbf{protocol} & \textbf{group} & \textbf{candidates} & \textbf{checked} & \textbf{runtime (s)} \\
    \hline
    \multirow{3}{*}{AXI} & handshake & 254 & 35 & 0.48 \\
                         & reset     & 240 & 50 & 0.38 \\
                         & low power & 240 & 37 & 0.77 \\
    \hline
    \multirow{2}{*}{Q-Channel} & handshake & 1182 & 218 &  8.07 \\
                               & reset     & 4096 & 850 & 37.43 \\
    \hline
    \multirow{2}{*}{P-Channel} & handshake & 1223 & 231 &  9.35 \\
                               & reset     & 4188 & 899 & 44.40 \\
    \hline
    \multirow{2}{*}{WISHBONE} & data transfer & 4464 & 484 & 19.71 \\
                              & reset         & 1182 & 185 &  2.32 \\
    \hline
    APB & data transfer & 4372 & 233 & 12.64 \\
    \hline
    \multirow{2}{*}{PCI} & transactions &  4280 &  823 & 16.66 \\
                         & termination  & 11250 & 2464 & 69.41 \\
    \hline
\end{tabular}
\end{table}

\subsection{LLM-assisted Property Filtering}
\label{sec:experiment:llm}

Given the formally checked property set and a textual description from the document, an LLM is then prompted to extract the properties from the formally checked set that best describe the details of the textual description.

Table~\ref{tab:llm} reports the results. The number of formally checked properties is shown in column~3, and the number of LLM-extracted properties is shown in column~4.
To analyze the quality of the extracted property sets, we perform a manual inspection and divide the extracted properties into several categories. The results are reported in the remaining columns:
\begin{itemize}
    \item \textbf{covered}: We analyze whether each target property is covered by the extracted property set. The target properties are further divided into \textit{explicit} and \textit{inferred} sets, where the explicit properties are obtained by analyzing the textual descriptions, while the inferred properties are not explicitly described in the textual descriptions but derived based on the user's understanding of communication protocols. 
    
    Note that there are multiple ways to describe a temporal constraint with SVA properties. For example, the semantics of a target property may be covered by two distinct properties from the LLM-extracted set. Therefore, when analyzing whether a target property is covered, we consider the full extracted set.
    \item \textbf{correct}: This category consists of the properties that are correct with respect to the protocol specifications. It can be further divided into \textit{unique} and \textit{redundant} sets. A property is unique if it enforces constraints on signal behavior that are not covered by any other properties. A property is redundant if removing it would not affect the semantics of the remaining unique set.
    \item \textbf{incorrect}: A property is categorized as incorrect if it violates the textual description of the specification document.
    For example, for the WISHBONE data transfer case, the LLM extracted a property stating that ``the acknowledgment signal can be asserted only when the request signal is asserted." However, the textual description allows the acknowledgment signal to be asserted at any time.
    Note that these incorrect properties are retained in the formally checked set when the timing diagrams capture only certain conditions, but not all conditions, that are allowed.
    \item \textbf{unsure}: Like a human engineer, the LLM may also infer properties that are not indicated in the textual descriptions. Some of them may require further clarification from the user. For example, the LLM inferred from the AXI low-power handshake mechanism that ``the acknowledgment signal should be asserted eventually in response to the request signal," which might hold true but is not mentioned in the specification document. We label these as \textit{unsure} properties.
    \item \textbf{hallucinated}: Finally, the LLM can also hallucinate, where it makes up properties that are not provided in the checked set, even though it is instructed to extract only properties from that set. These properties can be automatically removed from the extracted set.
\end{itemize}

\begin{table*}
\renewcommand{\arraystretch}{1.3}
\caption{Evaluation of LLM-based Property Filtering.}
\label{tab:llm}
\centering
\begin{tabular}{|c|c||c|c||c|c||c|c||c|c|c|}
    \hline
    \multirow{2}{*}{\textbf{protocol}} & \multirow{2}{*}{\textbf{group}} & \multirow{1}{*}{\textbf{checked}} & \multirow{1}{*}{\textbf{extracted}} & \multicolumn{2}{c||}{\textbf{covered}} & \multicolumn{2}{c||}{\textbf{correct}} & \multirow{2}{*}{\textbf{incorrect}} & \multirow{2}{*}{\textbf{unsure}} & \multirow{2}{*}{\textbf{hallucinated}} \\
    \cline{5-8}
    & & (by SAT) & (by LLM) & \textbf{explicit} & \textbf{inferred} & \textbf{unique} & \textbf{redundant} & & & \\
    \hline
    \multirow{3}{*}{AXI} & handshake & 35 & 3 & 2/2 & 0 & 2 & 0 & 0 & 1 & 0 \\
                         & reset     & 50 & 4 & 2/2 & 0 & 2 & 2 & 0 & 0 & 0 \\
                         & low power & 37 & 6 & 0 & 4/4 & 4 & 0 & 0 & 2 & 0 \\
    \hline
    \multirow{2}{*}{Q-Channel} & handshake & 218 & 14 & 11/11 & 0 & 11 & 3 & 0 & 0 & 0 \\
                               & reset     & 850 &  4 &   2/2 & 0 &  2 & 0 & 2 & 0 & 0 \\
    \hline
    \multirow{2}{*}{P-Channel} & handshake & 231 & 19 & 11/14 & 0 & 11 & 8 & 0 & 0 & 0 \\
                               & reset     & 899 &  4 &   2/2 & 0 &  2 & 2 & 0 & 0 & 0 \\
    \hline
    \multirow{2}{*}{WISHBONE} & data transfer & 484 & 14 & 1/2 & 5/5 & 6 & 4 & 4 & 0 & 0 \\
                              & reset         & 185 &  4 & 2/2 &   0 & 2 & 2 & 0 & 0 & 0 \\
    \hline
    APB & data transfer & 233 & 12 & 9/9 & 2/3 & 11 & 1 & 0 & 0 & 0 \\
    \hline
    \multirow{2}{*}{PCI} & transactions &  823 & 14 &  0 & 0/7 & 7 & 2 & 1 & 0 & 4 \\
                         & termination & 2464 & 20 & 6/20 & 0 & 10 & 2 & 1 & 1 & 0 \\
    \hline
\end{tabular}
\end{table*}

For most protocols (except PCI), the LLM-extracted properties successfully cover most (53/58) of the manually derived properties, including both explicit and inferred ones. This demonstrates the ability of LLMs to leverage ``common sense" for on-chip communication protocols, even when it is not explicitly explained by the textual descriptions.

We have found that LLMs (or at least the o1 reasoning model) tend to select redundant properties. Although this can make the LLM-extracted set larger and harder to read than a manually developed one, this might not be an issue for formal verification tools. For example, commercial model checkers can handle multiple properties together in a single verification run with multi-property verification engines. In this case, adding some redundant properties may not require extra proving effort from the model checker. However, further analysis with actual hardware and verification tools is required to evaluate the actual impact of these redundant properties.

Furthermore, the LLM-extracted properties are rarely incorrect. One possible reason is that the candidate properties have already been formally checked against the timing diagrams in the specification documents. Therefore, there weren't too many incorrect properties in the formally checked set to begin with.

\subsubsection{Discussion}

Compared to other protocols, we obtained much worse results for the PCI Local Bus specification, including the bus transactions (``transactions") and transaction termination (``termination") mechanisms. In the following, we discuss some reasons that affect the performance of our framework.

\textbf{Lack of textual descriptions.} To begin with, the PCI transactions case has the most complex data transfer mechanisms among all data transfer protocols in our experiments. We had to add two auxiliary signals to be able to write down the properties required for its functionality. (Note that it is common to use auxiliary signals when deriving complex temporal properties \cite{godhal2013synthesis}.) Apparently, none of these rules are directly indicated in the textual descriptions. We observed that the LLM still tries to infer some rules despite there being a lack of any described in the text. However, in this case, the inferred rules mostly regulate the behavior of auxiliary signals. Even though they are correct, none of them covers the target properties. In conclusion, experienced engineers are still more capable than LLMs in inferring properties when there is a lack of textual description, especially for complex cases.

\textbf{Limited reasoning capability of LLMs.} Even when a target property is explicitly described by the textual description, we have found that LLMs sometimes fail to reason about it. For example, the textual description of the PCI termination case contains the following information:
\small{
\begin{Verbatim}[breaklines=true, breaksymbolleft={}, commandchars=\\\{\}, fontsize=\small]
Note: A # symbol at the end of a signal name indicates that the asserted state occurs when the signal is at low voltage. When the # symbol is absent, the signal is asserted at a high voltage.
...
Once a master has asserted IRDY#, it cannot change IRDY# or FRAME# until the current data phase completes.
...
A data phase completes on any rising clock edge on which IRDY# is asserted and either STOP# or TRDY# is asserted.
\end{Verbatim}
}
From these descriptions, a human engineer can explicitly derive the following rules:
\begin{itemize}
    \item If IRDY\# is LOW and STOP\# is HIGH and TRDY\# is HIGH, then IRDY\# remains stable in the next cycle.
    \item If IRDY\# is LOW and STOP\# is HIGH and TRDY\# is HIGH, then FRAME\# remains stable in the next cycle.
\end{itemize}
However, the o1 reasoning model did not extract these two properties even though they were in the formally checked set. We conduct additional experiments to analyze factors that affect the model's reasoning capability.

\subsubsection{Additional Experiments}

We analyze two common factors -- \textit{input size} and \textit{text quality} -- that influence the reasoning capability of LLMs, by conducting experiments on the PCI termination case. We modify (1) the number of formally checked properties and (2) the textual description and observe whether the model is able to extract the two target properties shown in the example above.

Table~\ref{tab:analysis} reports the results of our analysis. We provide 3 levels of text quality: the original textual description (``original"), only relevant information extracted from the textual description (``extracted"), and directly the target properties (``direct"). We provide 4 different sizes of checked sets, where the smaller sets are created by randomly dropping properties from the original checked set.

\begin{table}
\renewcommand{\arraystretch}{1.3}
\caption{Quantitative analysis on the PCI transaction termination mechanisms. A checkmark (\cmark) indicates that the target properties are covered by the LLM-extracted property set. A cross (\xmark) indicates otherwise.}
\label{tab:analysis}
\centering
\begin{tabular}{|c|c|c|c|c|}
     \hline
     \diagbox{text quality}{checked set} & 2464 (original) & 500 & 100 & 20 \\
     \hline
     original & \xmark & \xmark & \xmark & \cmark \\
     \hline
     extracted & \xmark & \cmark$^{\mathrm{*}}$ & \cmark & \cmark \\
     \hline
     direct & \xmark & \xmark & \cmark & \cmark \\
     \hline
     \multicolumn{5}{l}{$^{\mathrm{*}}$Some incorrect properties were selected as well.}
\end{tabular}
\end{table}

As expected, the ability of the LLM to extract target properties degrades as the number of checked properties grows. It is also shown that the model generally performs worse when the textual description is more indirect and noisy.

Based on the analysis, we propose several methods that can possibly improve our framework. For example, for cases where the formally checked set still contains thousands of properties, a possible solution is to break the set into smaller subsets and perform a divide-and-conquer approach. For cases with poor text quality, we can possibly rewrite the textual description with LLM-based techniques. We will work on improving the results of these challenging scenarios in the future.

\subsection{Comparison of \sys with Previous Work} \label{sec:assertllm}

Previous research has proposed ML-based approaches to generate SVA properties from IC design specification documents for RTL design and verification purposes~\cite{parthasarathy2021spectosva, mali2024chiraag, yan2025assertllm}. We demonstrate the limitations of such approaches in generating formal specifications for on-chip communication protocols.

The AssertLLM framework \cite{yan2025assertllm} incorporates three customized LLMs that deal with decomposed tasks to generate SVAs from specification documents. Two of the LLMs are used to extract information from (1) textual descriptions, and (2) waveforms (timing diagrams) in the specification documents. The third LLM is used to translate the extracted information into SVAs. The generated SVAs are categorized into three types: bit-width, connectivity, and function. Since the first and second types are not relevant to signal interactions, we only include the function type SVAs in our results.

Table~\ref{tab:assertllm} evaluates the effectiveness of AssertLLM in generating SVAs for the Q-Channel protocol. The authors of AssertLLM used the OpenAI GPT-4o model~\cite{openai2024gpt4ocard} for their experiments when the paper was published. For fairness in comparison, we include results for AssertLLM with the more advanced o1 reasoning model, which is the same as what we used in our framework. Since many SVAs generated by the o1 model contain minor syntax errors that can be fixed straightforwardly (e.g., misusing $\neg$ instead of $\sim$ as the negation operator), we also include the results after fixing these errors.

\begin{table*}
\renewcommand{\arraystretch}{1.3}
\caption{Evaluating AssertLLM on the Q-Channel protocol.}
\label{tab:assertllm}
\centering
\begin{tabular}{|c|c|c|c||c|c|c|c|c|c|}
\hline
\textbf{Framework} & \textbf{model} & \textbf{total extracted} & \textbf{covered} & \textbf{syntax error} & \textbf{non-existing signals} & \textbf{incorrect} & \textbf{tautology} & \textbf{correct} \\
\hline
\multirow{3}{*}{AssertLLM} & GPT-4o & 38 & 2/13 & 16 & 11 & 4 & 4 & 3 \\
\cline{2-9}
                           & o1 & \multirow{2}{*}{16} & 5/13 & 9 & 0 & 1 & 2 & 4 \\
\cline{2-2} \cline{4-9}
                           & o1 + manual fixes & & 5/13 & 0 & 0 & 2 & 8 & 6 \\
\hline
\hline
\sys & o1 & 18 & 13/13 & 0 & 0 & 2 & 0 & 16 \\
\hline
\end{tabular}
\end{table*}

Q-Channel is a simple low-power interface, where all the rules required to construct its formal specification are clearly listed in its specification document~\cite{spec:lowpower}. However, even with the use of the entire document, AssertLLM was only able to cover a small portion of the target rules. This supports our claim that it is difficult using LLMs \emph{alone} to extract the correct information from protocol specification documents. Moreover, relying on LLMs to extract information increases the risk of generating problematic assertions, such as creating non-existing signals and generating tautologies (as reported in the table). Relying on LLMs to translate natural language into SVA also results in syntax errors.

\sys solves all these issues by generating necessary properties from a predefined grammar and automatically translating them into SVA. A formal approach is used to filter out properties inconsistent with timing diagrams, and the LLM is only responsible for extracting properties from the remaining set.

\section{Limitations} \label{sec:limitations}

\textbf{Timing diagrams.} \sys utilizes the unambiguous timing diagrams from the document in its property filtering stage to reduce incorrect properties. For simple protocols such as Q-Channel and P-Channel, all properties required to form the formal specification are described by timing diagrams. However, for more complex protocols, some mechanisms that regulate signal behaviors are merely described in text, possibly complemented with other contexts, such as tables and flowcharts. In our experiments, we focus on the functionality groups that utilize descriptive timing diagrams. Therefore, for more complex protocols (AXI, WISHBONE, APB and PCI), the target properties in our experiments do not cover the entire formal specification.

One alternative is to skip the SAT-based property check and directly extract the necessary properties using LLM-based techniques. However, the performance of the LLM can be hindered if the candidate property set becomes too large (as shown in Section~\ref{sec:experiment:llm}). Another solution is to let users create their own timing diagrams. This can be beneficial when the user is not familiar with SVA but understands how the protocol works.

\textbf{Textual descriptions.} The LLM-based property filtering matches the formally checked property set with the textual description to retain only necessary properties. However, as discussed in Section~\ref{sec:experiment}, some properties are not described by text and are inferred based on the user's general understanding of communication protocols. In Section~\ref{sec:experiment:llm}, we have observed that LLMs do have some capability in inferring properties, but are not as good as an experienced engineer. It is also shown that text quality affects LLMs' reasoning ability.

In our experiments, we tried not to modify the way the texts were written when extracting them from the documents. However, note that users can provide their own textual descriptions, possibly by modifying the original documents and adding more information.

In summary, \sys aims to assist, but not replace, its users in deriving formal specifications for on-chip communication protocols.

\section{Related Work} \label{sec:related}

\textbf{Sentence-level NL to SVA translation.} Numerous studies have worked on translating natural language sentences to SVA properties. These sentences are provided in a format that directly describes temporal behaviors of several signals, possibly obtained by analyzing design or protocol specification documents. Natural language processing (NLP) techniques are then used to translate these sentences into SVA. This includes more traditional grammar- and rule-based methods \cite{soeken2014automating, ghosh2016arsenal, harris2013capturing, harris2015generating, harris2016glast, krishnamurthy2019controlled, zhao2019automatic, krishnamurthy2019ease, frederiksen2020automated} and emerging ML technologies such as LLMs \cite{keszocze2019chatbot, parthasarathy2021spectosva, aditi2022hybrid, cosler2023nl2spec, sun2023towards, kande2023llm, liu2024domain, yan2025assertllm, quddus2024enhanced}.

\textbf{Information extraction from specification documents.} In addition to sentence-level SVA translation, several works have addressed the problem of extracting property-related information from specification documents. SpecToSVA \cite{parthasarathy2021spectosva} developed an ML classifier to identify sentences that correspond to functional properties from IC design specifications, before translating them into SVA.
ChIRRAG \cite{mali2024chiraag} manually modifies design specifications into a specific format with distinctive labels, and prompts an LLM to generate SVA properties directly from the modified specifications. The generated assertions are further verified in a simulation environment, and the prompt will be manually and repeatedly refined based on the errors. AssertLLM \cite{yan2025assertllm} incorporates multiple LLMs to deal with decomposed tasks, including extracting structured information from documents and translating them into SVAs.

These approaches primarily focus on generating SVA properties for RTL debugging. Our evaluation of AssertLLM in Section~\ref{sec:assertllm} demonstrated that LLMs are poor at precisely extracting the required information from protocol specification documents. Furthermore, LLMs can generate problematic SVAs, even including syntax errors. \sys solves these issues by combining formal approaches with LLM-based techniques.

\textbf{Generating assertions from other sources.} Previous works have also considered other sources from which SVA or other forms of temporal assertion can be generated. For example, numerous works have considered generating temporal assertions from hardware simulation traces with static analysis and data mining approaches \cite{li2010scalable, vasudevan2010goldmine, liu2011automatic, liu2012word, hertz2013mining, sheridan2014coverage, danese2017team, germiniani2022harm}. Other approaches study the structures of SVA properties for specific applications to construct specifications or ``property libraries." Given an RTL design of the application, SVAs can then be automatically generated based on these predefined specifications \cite{manerkar2017rtlcheck, orenes2021autosva}. More recently, researchers have also considered the use of LLMs to generate SVAs directly from RTL source codes \cite{orenes2023using, pulavarthi2024assertionbench}. These works are orthogonal to ours and could be used in combination with the \sys framework.

\section{Conclusions and Future Work} \label{sec:conclusion}

We propose \sys, a two-stage framework that combines formal and LLM-based techniques to help hardware engineers construct formal protocol specifications from informal documents. We evaluated the effectiveness of \sys for generating SVA properties from specification documents of various open-source on-chip communication protocols, and demonstrated that it outperforms a state-of-the-art SVA generation framework. In addition, we analyze cases where \sys missed many target properties and identify factors that affect its performance.

Future work will focus on improving the algorithms to handle these challenging cases, following the directions pointed out in Section~\ref{sec:experiment:llm}. In general, it is also worth exploring more advanced LLM technologies, such as few-shot learning and fine-tuning.

\bibliographystyle{IEEEtran}
\bibliography{references}

\end{document}